\documentclass[12pt]{article}

\usepackage{newtxtext,newtxmath}
\usepackage{graphicx}

\usepackage{siunitx,xfrac}
\usepackage{bm}

\usepackage[letterpaper,margin=1in]{geometry}

\renewenvironment{abstract}
	{\quotation}
	{\endquotation}
\date{}

\makeatletter
\renewcommand{\fnum@figure}{\textbf{Figure \thefigure}}
\renewcommand{\fnum@table}{\textbf{Table \thetable}}
\makeatother

\usepackage{url}

\def\scititle{
	Morpho-plastic cellular metamaterials
}
\title{\bfseries \boldmath \scititle}

\author{
Victor Charpentier$^{1}$, Ignacio Andrade-Silva$^{1,2}$, Trevor J. Jones$^{3,4}$, Tom Marzin$^{3},$\\ Stéphane Bourgeois$^{5}$, P.-T. Brun$^{3}$, Joel Marthelot$^{1\ast}$
\\
\small{$^{1}$Aix-Marseille Univ, CNRS, IUSTI, Marseille, France}\\
\small{$^{2}$Departamento de Física, Facultad de Ciencias Físicas y Matemáticas, Universidad de Chile, Santiago, Chile}\\
\small{$^{3}$Department of Chemical and Biological Engineering, Princeton University, Princeton, NJ 08540, USA}\\
\small{$^{4}$Department of Mechanical Engineering, Carnegie Mellon University, Pittsburgh, PA, USA}\\
\small{$^{5}$Aix-Marseille Univ, CNRS,} \small{Centrale Méditerranée, LMA, Marseille, France}\\
\small{$^\ast$Corresponding author. Email: joel.marthelot@univ-amu.fr}
}

\begin{document} 

\maketitle

\begin{abstract} \bfseries \boldmath
Deployable structures, essential across various engineering applications ranging from umbrellas to satellites, are evolving to include soft, morphable designs where geometry drives transformation. However, a major challenge for soft materials lies in achieving reliable actuation and stable shape retention in their deployed state. Drawing inspiration from biological growth processes, we demonstrate that irreversible plastic deformations can be leveraged to create cellular metamaterials with permanent morphing capabilities. By employing a simple actuation method, stretching and releasing the structure's ends, our approach facilitates the design of structures capable of sequential, multi-target configurations and mechanical multistability. Our methodology augments additive manufacturing to transform flat, printable designs into intricate 3D forms, with broad applications in consumer goods, healthcare, and architecture.
\end{abstract}

\newpage

\begin{figure}
  \centering
 \includegraphics[width=\textwidth]{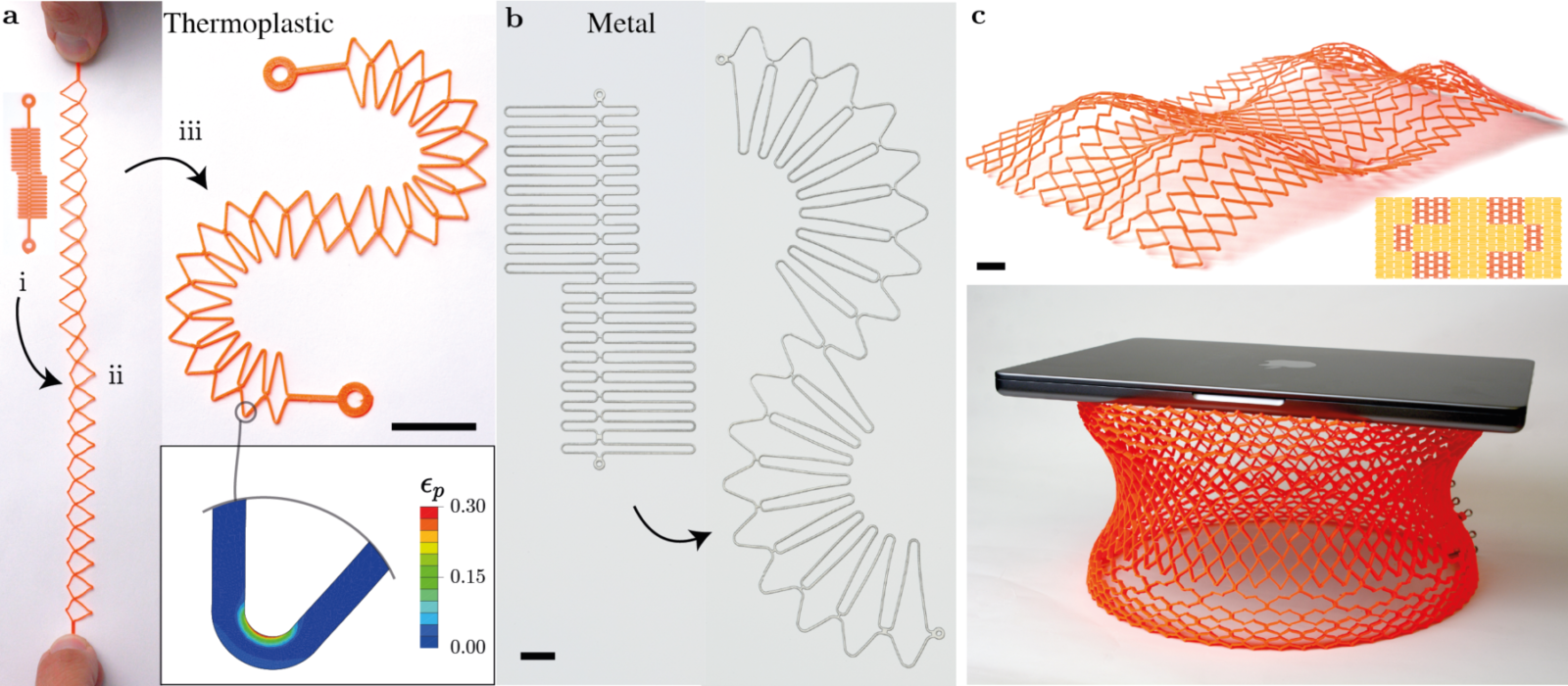}
  \caption{\textbf{Shaping with plasticity}. 
  (\textbf{a}) (i) A 3D-printed chain of cells, (ii) stretched by a factor $\lambda=5.5$, (iii) adopting a resting shape with the desired target curvature. (Inset) Finite element model showing the localization of plastic deformation in the hinges. Color encodes the equivalent plastic strain $\varepsilon_p$.
  (\textbf{b}) Titanium cell chain and its resting shape after stretching by a factor $\lambda=2.5$.  (\textbf{c}) Generalization of the strategy to program plates with positive and negative Gauss curvatures, and vase shapes with load-bearing capabilities (13" laptop, 1.6 kg mass). Scale bars 20 mm.}
  \label{fig:0}
\end{figure}

Irreversible deformations have long been used to create complex shapes and structures across various materials, including ceramics, metals or polymers. Examples range from traditional pottery and forging to industrial casting, molding and additive manufacturing. In these processes, the material is directly shaped into the desired form using a series of applied constraints coupled to solidification, e.g., via a chemical reaction like curing in elastomers, thermoplastic or concrete.  Dramatic shape transformations can also be achieved through the elastic deformation of flexible materials in response to external forces or fields\,\cite{bertoldi2017flexible}. Inflatable systems can take on intricate shapes by programming 1D curvilinear paths\,\cite{siefert2019programming,jones2021bubble,baines2023programming,andrade2023fabric}, while flat 2D structures can be transformed into 3D target surfaces via cuts\,\cite{pikul2017stretchable, celli2018shape, choi2019programming, jin2020kirigami, hong2022boundary, mcmahan2022effective}, folds\,\cite{felton2014method,rus2018design,dudte2016programming,dieleman2020jigsaw, melancon2021multistable,melancon2022inflatable}, bilayers~\cite{brau2011multiple,ramachandran2021uniaxial,zhang2022shape}, networks of beams and hinges\,\cite{baek2018form, panetta2019x, guseinov2020programming,jones2023soft}; or using a change of metric through inflation~\cite{siefert2019bio} or swelling\,\cite{klein2007shaping,kim2012designing,sydney2016biomimetic,aharoni2018universal}. While large-scale structures have been realized\,\cite{filipov2015origami,melancon2021multistable}, most applications in soft robotics\,\cite{hwang2022shape,sedal2020design} and soft electronics\,\cite{xu2015assembly} are typically not structurally functional due to the trade-off between load capacity and compliance. Further, they do not retain their shape in the deployed state, thereby requiring active actuation for practical use. 

\begin{figure}
  \centering
 \includegraphics[width=1\textwidth]{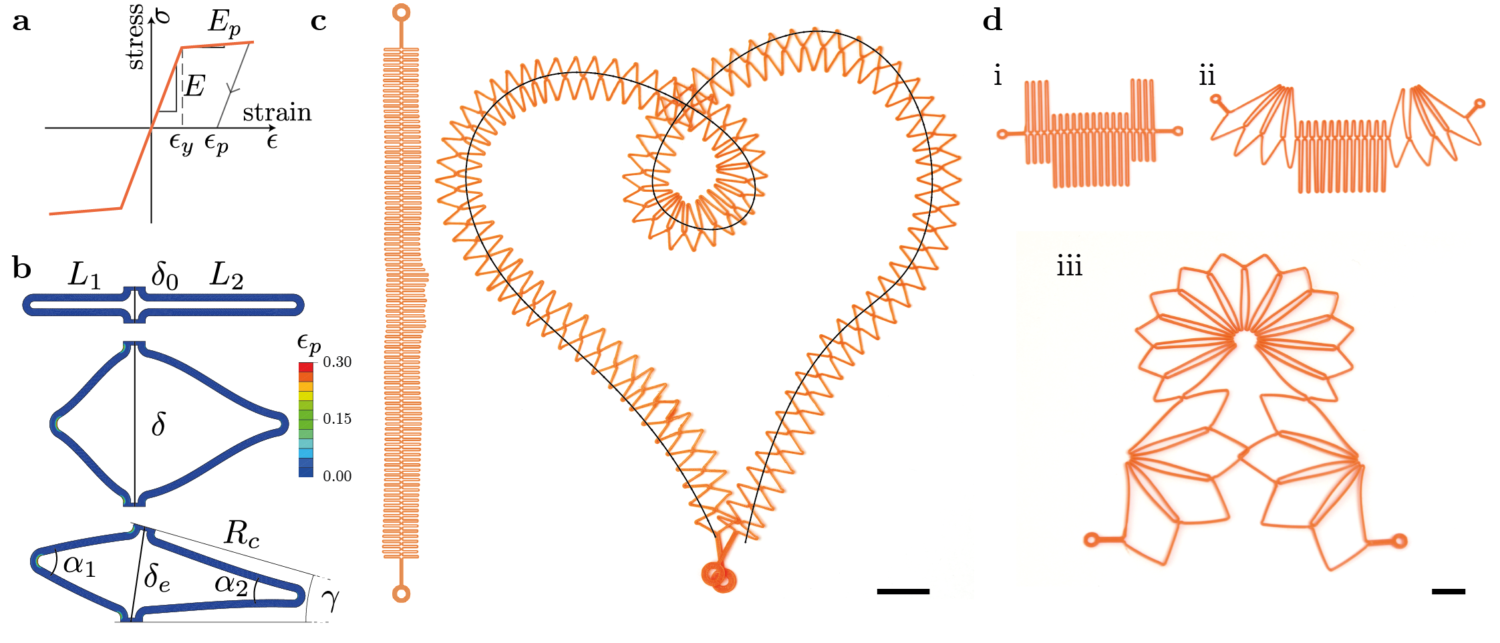}
  \caption{\textbf{Programmed planar shape}. (\textbf{a}) Elasto-plastic model for the material. (\textbf{b}) Finite elements predictions of cell shapes during extension ($\lambda=\delta/\delta_0=5.5$) and relaxation. Color encodes the magnitude of equivalent plastic strain $\varepsilon_p$. (\textbf{c}) Self-intersecting complex shape and target trajectory (solid black line). (\textbf{d}) Multiple shapes programmed by applying different maximum stretch factors: (i) $\lambda=1$, (ii) $\lambda=4.7$, (iii) $\lambda=8.3$. Scale bars 10 mm.}
  \label{fig:1}
\end{figure}

Here, inspired by natural morphogenesis, where differential tissue growth induces shape changes\,\cite{sharon2002buckling, liang2011growth, ben2013anisotropic}, we introduce cellular metastructures that can morph into structurally functional and programmable shapes. These structures are designed to transform simply by pulling and releasing their boundaries (see Fig.\,\ref{fig:0}a), enabling permanent morphing in both 2D and 3D configurations. The retention of these shapes is achieved through localized plastic deformations in specific areas where the yield stress is exceeded. Thus, our strategy is versatile and applicable to any material with elasto-plastic properties, including thermoplastics (Fig.\,\ref{fig:0}a) and metals (Fig.\,\ref{fig:0}b). Building on our understanding of how simple cellular chains morph along curvilinear paths, we tackle the inverse design problem, advancing our approach by encoding geometric incompatibilities into 2D flat surfaces that morph into 3D shells (see Fig.\,\ref{fig:0}c) and opening up new possibilities for engineering applications. 

Fig.\,\ref{fig:0}a illustrates our approach with a 3D-printed chain of cells. When the chain is stretched, plastic deformations are localized in the hinges (see the finite elements simulations in Inset). Upon released, the chain recoils due to springback and adopts a curved shape at equilibrium, which can be programmed via the asymmetry of the cells in the initial configuration. To explain how cell asymmetry controls shape morphing, we define the material and geometric properties of a single cell prior to, during and post loading. The stress-strain behavior is modeled as a linear strain hardening elasto-plastic material characterized by Young modulus $E$, yield strain $\varepsilon_Y$, and tangent modulus $E_p$ (see Fig.~\ref{fig:1}a and Supplementary Materials, `Material behavior'). Each cell consists of four straight beams (arms) connected by curved segments (hinges), with asymmetric arm lengths $L_1$ and $L_2$ (Fig.~\ref{fig:1}b). The chain is stretched by a factor $\lambda=\delta/\delta_0$, inducing plastic deformation beyond the yield strain. Upon load release, plastic deformation results in a new length $\delta_e$ and residual opening angles $\alpha_1$ and $\alpha_2$ differ, causing the unit cell to tilt with angle $\gamma$. 

\begin{figure}
  \centering
  \includegraphics[width=\textwidth]{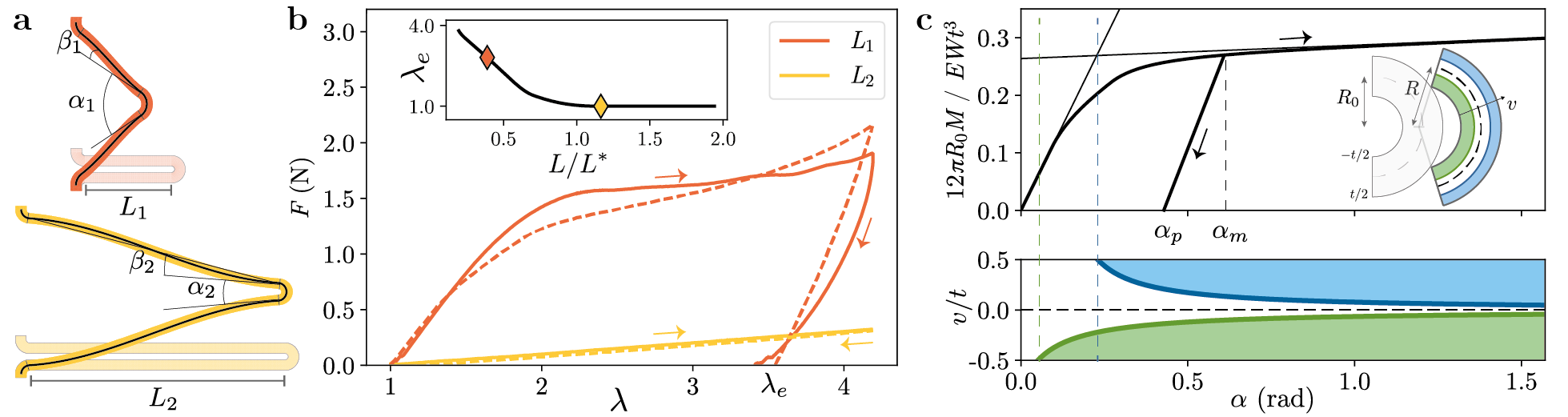}
  \caption{\textbf{Mechanics of half-cells}. (\textbf{a}) Finite element predictions for $\lambda=4.2$ of the deformed configuration of cells with two different arm lengths $L_1=4$\,mm (red) and $L_2=12$\,mm (yellow), beam thickness $t= 0.51$ mm and initial radius $R_0= 0.4$\, mm. Our model prediction (solid black line) is superimposed on the finite element results. (\textbf{b}) Force-displacement curves for the two different geometries. Experiments are represented by solid lines and model predictions by dotted lines. (Inset) Equilibrium stretch $\lambda_e$ for a maximum stretch $\lambda=4.2$ plotted against arm length. (\textbf{c}) Prediction of the hinge moment and plasticity in the hinge cross-section plotted against opening angle $\alpha$. (inset) Model of the elasto-plastic deformation of the hinge for angle $\alpha=0.6$.}
    \label{fig:2}
\end{figure}

Cell deformation is predicted by finite element simulations. We find that the beams act as lever arms, concentrating the bending moment in the hinges during extension. Crucially, the hinge connected to the shorter arm accumulates more plastic deformation, leading to a larger opening angle at equilibrium. This asymmetry has two origins: (i) basic geometry enforces that the opening angle of the shorter arms $\alpha_1$ is greater than that of the longer arms $\alpha_2$; and (ii) longer arms bend more, which induces additional plastic deformation in shorter hinges, further amplifying the disparity between the two. 

Arbitrarily complex shapes in the plane can be achieved by assembling cells (see Fig.~\ref{fig:1}c). The cell array is generated by discretizing the curvature of the target trajectory into elements with identical $\delta_e$ and discrete radius of curvature $R_c\sim\delta_e/\gamma$ to obtain the target shape. Note that self-intersections typically do not hinder programmability (see Fig.~\ref{fig:1}c). Additionally, multiple shapes can be programmed by incrementally increasing the applied stretch (see Fig.~\ref{fig:1}d). With these design principles, we can solve the inverse problem using the following model. 

We first consider the deformation of half-cells, characterized by the hinge angle $\alpha$ and the bending angle $\beta$  (Fig.\,\ref{fig:2}a). Fig.\,\ref{fig:2}b illustrates the mechanical response of a chain of symmetric cells under loading ($\lambda=4.2$). In experiments, the force recorded for the short arm ($L_1=4$\,mm, solid red line) initially increases linearly with displacement before reaching a plateau. Upon unloading, the cell exhibits permanent deformation with a residual stretch $\lambda_e$. In contrast, cells with longer arms ($L_2=12$\,mm, solid yellow line) remain within the elastic regime under the same displacement, showing no permanent deformation.

\begin{figure}
  \centering
  \includegraphics[width=1\textwidth]{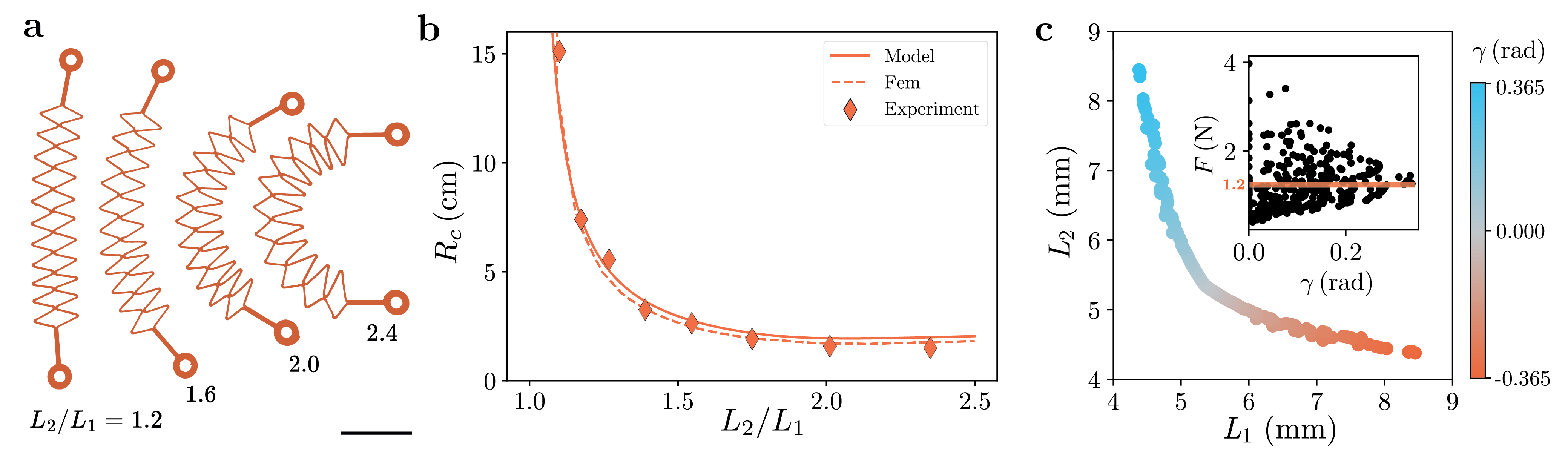}
  \caption{\textbf{Asymmetric cells}. (\textbf{a}) Chain of asymmetric cells with different ratio $L_2/L_1$ after stretching by $\lambda=4.2$. (\textbf{b}) Radius of curvature of the deformed chain plotted against arm lengths $L_2/L_1$ for a constant $\lambda=4.2$. (\textbf{c}) Tilt angle $\gamma$ plotted against geometry with constant force $F=1.2$\,N and uniform length $\delta_e=7.5$\,mm. (Inset) Span of the tilt angle $\gamma$ at constant force. Scale bar 20\,mm.}
  \label{fig:3}
\end{figure}

To further predict the mechanical response of the cell, we consider elasto-plastic deformations in the hinges and purely elastic deformations in the beams connecting them. We assume a uniaxial stress state, with the hinges remaining circular during deformation (Inset of Fig.\,\ref{fig:2}c), allowing for an analytical description of the strain across the thickness of the central hinge (see Supplementary Materials, `Hinge model'). As the hinge opens, plasticity initially emerges below the neutral line in regions under tension (shown in green in Fig.\,\ref{fig:2}c). With increasing $\alpha$, plastic deformations also develop above the neutral line in regions under compression (shown in blue in Fig.\,\ref{fig:2}c). The hinge moment is $M(\alpha) = \int^{t/2}_{-t/2} v \, \sigma(\varepsilon(v,\alpha)) dv$, where $t$ is the thickness, $v$ the normal offset from centerline and $\sigma$ the stress. The hinge moment is calculated analytically in the three regimes indicated by the vertical dotted lines in Fig.\,\ref{fig:2}c: fully elastic, with a single tensile plasticized region and with two plasticized regions. The system behaves as a piecewise linear torsional spring, as indicated by the change in slope in Fig.\,\ref{fig:2}c. When the cell is released, the hinge angle relaxes elastically from its maximum value $\alpha_m$ to a new equilibrium position defined by $\alpha_{p}$, following the slope of the elastic regime. The initially straight beams do not undergo plastic deformation, so their bending energy depends solely on their flexural rigidity and the angle $\beta$ (see Supplementary Materials, `Arm bending'). Minimizing the energy of the system leads to an implicit kinematic relationship between $\alpha$ and $\beta$ (see Supplementary Materials, `Lagrangian formulation for symmetric cells'). 

To validate our model, we superimpose the shape predicted by the analytical model (solid black line) onto the numerical calculations in Fig.\,\ref{fig:2}a, finding favorable agreement. The model also predicts the force as a function of displacement, shown as a dashed line in Fig.\,\ref{fig:2}b, which aligns well with experiments. In the inset of Fig.\,\ref{fig:2}b, we show the predicted residual stretch $\lambda_e$ as a function of arm lengths. We observe a transition from permanent deformation in cells with shorter arms to an elastic response in those with longer arms. This transition occurs at a critical length $L^*$, which depends on $\lambda$, yield strain $\varepsilon_Y$, and cell geometry (see Inset in Fig.\,\ref{fig:2}b and Supplementary Materials, `Origami length'). 
We now turn to rationalizing the mechanical response of asymmetric cells, composed of two different arm lengths $L_1$ and $L_2$. When the cell is pulled and released, it adopts a tilted configuration (see Fig.\,\ref{fig:3}a) defined by an angle $\gamma$, an equilibrium joint-to-joint distance $\delta_e$, and a discrete radius of curvature $R_c = \delta_e/(2\sin(\gamma/2))$.
To predict $R_c$, we minimize the energy of the asymmetric cell (see Supplementary Materials, `Lagrangian formulation for asymmetric cells'). Our prediction for $R_c$ is shown in Fig.~\ref{fig:3}b, compared with experiments and full-fledged simulations for a fixed $\lambda=4.2$. 

When pulling a chain of cells, all cells are subjected to the same force $F$, as gravitational effects are negligible. This simplification enables us to solve the inverse problem of determining the initial cell geometry required to transform the chain into an arbitrary target curve. We first discretize the target trajectory into straight segments of uniform length $\delta_e$ and tilt angles $\gamma$. We then select the value of actuation force that maximizes the range of $\gamma$ (see Fig.~\ref{fig:3}c (Inset)). Using a multivariate linear interpolation routine, we create a mapping $\{F,\delta_e,\gamma \}\rightarrow \{L_1,L_2 \}$. Having chosen $F$ and $\delta_e$, our interpolated function provides the pair $\{L_1,L_2 \}$ that achieves the local target $\gamma$. As shown in Fig.~\ref{fig:1}c, this method yields strong agreement between our target and experimental results. The chain deployment is a dynamic process, not modeled here, which still converges to the target shape even when the chains self-intersect. Note, however, the need to operate at a prescribed force, such that this deployment type is best achieved with a force sensor. Alternatively, the force can be translated into a maximum displacement, enabling manual deployment. To further enhance the practicality of our methodology, we now demonstrate its generalization to 3D shapes.  

\begin{figure} 
  \centering
  \includegraphics[width=1\textwidth]{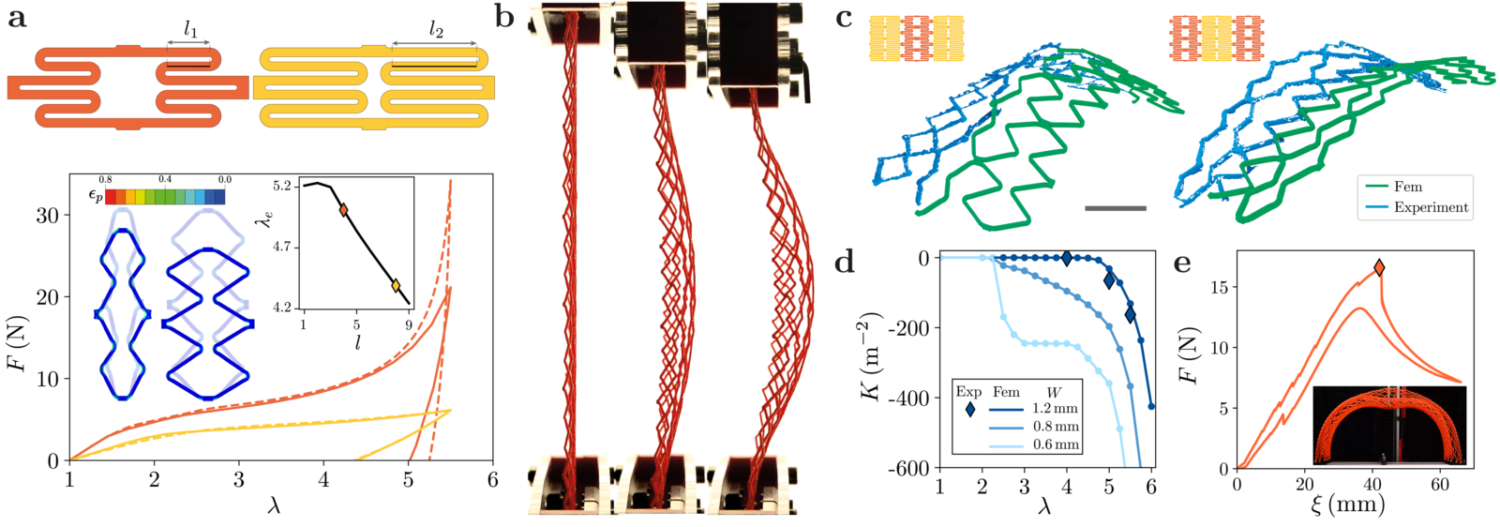}
  \caption{\textbf{Programmed out-of plane shape.} (\textbf{a}) Cells with internal length $l_1=4$\,mm (red) and $l_2=8$\,mm (yellow). Finite element prediction for a maximum stretch  of $\lambda=5.5$ and force-stretch curve. Dotted lines correspond to Kirchhoff rod models, solid lines to finite element results (Inset) Equilibrium cell stretch $\lambda_e$ plotted against $l$. (\textbf{b}) Relaxation of the plate, from planar at maximum stretch to out-of-plane at equilibrium.  (\textbf{c}) Positive (left) and negative (right) Gauss curvature out-of-plane deformation for 3d-scanned experiments and FEM with three parallel cell chains at $\lambda=5.5$ (\textbf{d}) Numerical and experimental Gauss curvature as function of $\lambda$ for three-parallel chains structure of (c) (\textbf{e}) A large assembly of cells creating a out-of-plane arched shape presents load-bearing capabilities. Scale bars 20 mm.}
  \label{fig:4}
\end{figure}

Our approach generates 3D shapes when multiple parallel chains are linked together (Fig.\,\ref{fig:0}c). To program the local curvature of the sheets, we adapt the base cell design, tessellating the plane with uniformly sized rectangular pixels. Each cell comprises 12 arms connected by hinges, with constant initial arm widths and hinge radii. The internal arm length $l$ is adjustable (Fig.~\ref{fig:4}a). After deformation, cells with shorter internal arms ($l_1=4$ mm shown in red) exhibit greater residual equilibrium heights compared to cells with longer internal arms ($l_2= 8$ mm shown in yellow), as shown by the finite element calculations. Our previous model for in-plane deformation of single chains assumes plastic deformations are confined to the hinges alone. However, with parallel-chain designs, we find irreversible deformations in the beams as well. We therefore introduce a general Kirchhoff rod centerline model in which plastic deformation can occur throughout the beam (see Supplementary Materials, ‘Kirchhoff rod model’). This model, in line with finite elements predictions (Fig.~\ref{fig:4}a), links deformation to the beam's rest curvature, which updates during plastic deformation. By adjusting the internal arm lengths $l$, we can program the equilibrium cell stretch $\lambda_e$ after release (Inset of Fig.~\ref{fig:4}a).

These cells tile the plane and therefore have no intrinsic curvature. However, geometric incompatibilities arise when connected cells with different internal arm lengths $l$ are plastically deformed, leading to out-of-plane deformation. The boundaries of the initial grid structure are fixed to custom-built experimental clamps, featuring sliders that allow free movement perpendicular to the tensile axis\,\cite{liu2022triclinic} (see Supplementary Materials, `Experimental clamps'). Once the boundaries are released after unloading, the structure deforms with significant Gauss curvature (see Fig.\,\ref{fig:4}b). The emergence of curvature is associated with changes in plate metrics, akin to in-plane strain gradients\,\cite{sharon2010mechanics}. By adjusting the internal arm lengths $l$, we control the stretch factor after release $\lambda_e$, thereby programming the curvature. Minimal grids for observing positive (dome) and negative (saddle) Gauss curvature consist of three parallel chains, as shown in Fig.~\ref{fig:4}c. Experimental shapes are 3D scanned and shown in blue, and compared with direct finite element simulations. The Gauss curvature $K$, defined as the product of the principle curvatures, is extracted from the 3D scans and compared with finite element predictions in Fig.~\ref{fig:4}d. The Gauss curvature emerges above a critical stretch and increases with the magnitude of the imposed displacement. For thinner plates ($W=0.8$\,mm), the critical stretch required to induce out-of-plane deformations decreases, and the magnitude of the curvature increases. For even thinner plates ($W=0.6$\,mm), out-of-plane deformations appear at the hinges appear during extension (see Supplementary Materials, `Control of curvature in 2D and 3D'), resembling the behavior of kirigami\,\cite{dias2017kirigami}. These local deformations do not hinder the metric difference between adjacent cells, allowing the plate to deform globally out-of-plane after unloading. The resulting 3D shapes exhibit load-bearing capacities, capable of supporting objects weighing several kilograms (Fig.\,\ref{fig:0}c). To assess the structure's rigidity, we subject an actuated shell to point loading at the apex (Fig.~\ref{fig:4}e). We find a typical load-to-self-weight ratio of 20, higher than existing plasticity-based morphing systems\,\cite{hwang2022shape}.  

In classical plate theory, the bending energy is negligible compared to the energy associated with in-plane strain gradients for sufficiently thin plates. As such, classical plate shapes correspond to stretch-free configurations that conform to the target metric~\cite{sharon2010mechanics}. In our case, the bending and stretching energies of the plate are of same order of magnitude. As a result, the target metric serves as an estimate for the resulting Gauss curvature, providing guidelines for the inverse design process. By strategically distributing discrete patches of positive and negative Gauss curvature across the plane, we demonstrate the ability to create complex shapes (Fig.\,\ref{fig:0}c). The height profile of these complex structures can be precisely controlled by adjusting the cell geometry, patch size, and the maximum stretch (see Supplementary Materials Fig.\,\ref{fig:kirigami}b). Longer samples with patches of curvature interconnected by regions of zero curvature result in the formation of multistable structures (see Supplementary Materials Fig.\,\ref{fig:Methods6}). These structures can be reconfigured by changing the state (direction) of a local feature, such as a doubly curved patch, through the application of an external force, a characteristic shared with other folded structures\,\cite{silverberg2014using, meeussen2023multistable}.

In conclusion, morpho-plastic cellular metamaterials leverage intelligent patterning combined with standard loadings to create intricate structures. This approach emphasizes design and mechanics over direct forming, advancing the field of mechanical metamaterials by embrassing plasticity, transitioning its use from shock absorption\,\cite{liu2024harnessing} to shape morphing. By transforming easily fabricated flat structures into complex, overlapping topologies and 3D geometries, this method simplifies the manufacturing of complex structures to the level of producing flat counterparts. The proposed tensile loadings are straightforward to implement across scales, from submicron to meter, with the resulting transformations persisting after actuation. Moreover, a single structure can be reconfigured multiple times into multiple forms, opening doors to a wide range of applications. The concept of using elasto-plastic behavior coupled with standard loadings to manufacture structures is versatile, adaptable to various materials and systems, all while maintaining a tractable modeling complexity.

\newpage

\renewcommand{\thefigure}{S\arabic{figure}}
\renewcommand{\thetable}{S\arabic{table}}
\renewcommand{\theequation}{S\arabic{equation}}
\renewcommand{\thepage}{S\arabic{page}}
\setcounter{figure}{0}
\setcounter{table}{0}
\setcounter{equation}{0}
\setcounter{page}{1}

\begin{center}
\section*{Supplementary Materials for\\ \scititle}

Victor Charpentier, Ignacio Andrade-Silva, Trevor J. Jones, Tom Marzin, Stéphane Bourgeois, P.-T. Brun, Joel Marthelot$^\ast$\\ 
\end{center}

\subsection*{Materials and Methods}

\subsubsection*{Sample fabrication} 

\noindent Polyethylene terephthalate glycol-modified (PETG) is printed using a Prusa MK3s+ 3D printer, with a 0.2 mm nozzle for curvilinear path assemblies (chains) and a 0.4 mm nozzle for out-of-plane assemblies (plates). The filament used is marine blue and orange Prusament PETG with an initial diameter of 1.75\,mm. The beam and hinge thicknesses are set to $t= 0.5$\,mm for chains and $t= 0.85$ mm for plates. The depth of the bar-hinge structures is maintained at $W=0.6$\,mm for chains and $W =1.2$\,mm for plates. For titanium chains, cells are water-jet cut from a 1.5\,mm thick grade 5 titanium plate, with beam and hinge thicknesses set to $t=1$\,mm.

\vspace{8mm}

\subsubsection*{Experimental clamps}

\begin{figure}
  \centering
  \includegraphics[width=.8\textwidth]{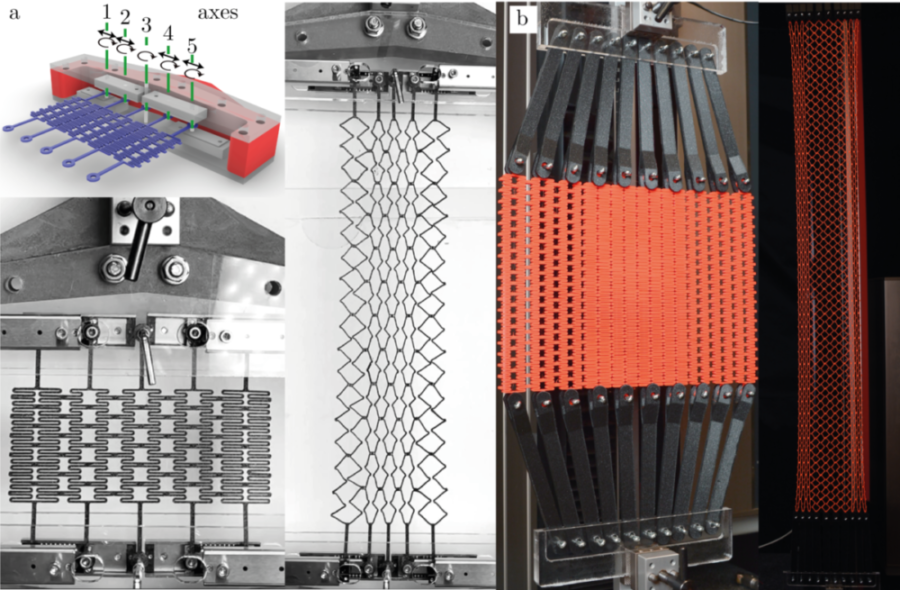}
  \caption{ \textbf{Experimental clamps.}
  (\textbf{a}) Sample clamps with sliders. Five-axis clamps for testing samples with five parallel chains. Five chains of six cells each at initial and maximum stretch. Plates in initial position (left) and at maximum stretch (right). (\textbf{b}) Sample clamps with pivoting connection bars. Up to 10 parallel chains can be tested. Plates in initial position (left) and at maximum stretch (right).
  }
  \label{fig:Methods2}
\end{figure}

\noindent We perform experimental uniaxial tension tests to characterize assemblies with varying numbers of parallel chains. Depending on the sample size, two different testing systems are employed. Samples with up to five parallel chains are tested using slider-equipped clamps. To minimize boundary effects, we introduce a new experimental setup that permits transverse displacements at the boundary (see Fig.\,\ref{fig:Methods2}). We use compact linear motion rolling guides with 32 mm travel (ball bearing sliders IKO BSR1550SL). Axes 1, 2, 4, and 5 are mounted on independent sliders, allowing for both rotation and translation (see Fig.\,\ref{fig:Methods2}a). The central axis is restricted to rotation only, preventing rigid translations of the entire structure. This experimental setup accommodates non-uniform patterns, as each connection can move independently of its neighbors. At maximum stretch, the distance between chains varies, but the chains remain vertical, demonstrating effective mitigation of boundary effects. This setup enables the testing of small samples with strong agreements between experimental results, analytical predictions and numerical models. For samples with more than five parallel chains, pivoting connection bars are used (see Fig.\ref{fig:Methods2}b), enabling tests of up to 10 parallel chains. These connection bars (in black) are secured to the testing machine on a rigid, evenly perforated frame, and are connected to the chains (in orange) with pins. The spacing of the holes is designed to match the average cell width at full extension. Although some boundary effects are visible at maximum extension, their impact on the equilibrium structure remains minimal.  

\vspace{8mm}

\subsubsection*{Material behavior} 

\noindent We consider a bilinear constitutive model to describe the stress softening behavior of elasto-plastic materials, defined as:
\begin{equation}
    \sigma(\varepsilon) = \begin{cases}
        E_p (\varepsilon+\varepsilon_Y) - E \varepsilon_Y & \varepsilon \le -\varepsilon_Y, \\
        E \varepsilon & -\varepsilon_Y < \varepsilon < \varepsilon_Y, \\
        E_p (\varepsilon-\varepsilon_Y) + E \varepsilon_Y & \varepsilon \ge \varepsilon_Y.
    \end{cases}
\end{equation}
where $\varepsilon_Y$ is the yield strain, $E$ is the Young's modulus, and $E_p$ is the plastic modulus following yielding. Upon stress release, we model plasticity assuming that stress is relieved following the linear Young's modulus, making the history-dependent stress a function of the maximum strain $\varepsilon_{max}$:
\begin{equation}
    \sigma(\varepsilon) = \begin{cases}
        E(\varepsilon - \varepsilon_{max}) + E_p \varepsilon_{max} & \varepsilon_{max} \le -\varepsilon_Y, \\
        E \varepsilon & -\varepsilon_Y < \varepsilon_{max} < \varepsilon_Y, \\
        E(\varepsilon - \varepsilon_{max}) + E_p (\varepsilon_{max}-\varepsilon_Y) + E_p \varepsilon_Y & \varepsilon_{max} \ge \varepsilon_Y.
    \end{cases}
    \label{eqn:bilinear}
\end{equation}

The material parameters for PETG are determined through tensile tests conducted on the extruded material. While PETG undergoes both striction and drawing, this elasto-plastic model provides an approximate representation of its mechanical response. The identified parameters are: Young's modulus $E=1.57$\,GPa, plastic modulus $E_p=30.8$\,MPa, yield strain $\varepsilon_{y}=\num{2.85e-2}$ and ultimate strain $\varepsilon_{u}=\num{2.42}$. For grade 5 titanium, we use the following parameters: Young's modulus $E=100.3$\,GPa, plastic modulus $E_p=0.79$\,GPa, yield strain $\varepsilon_{y}=\num{8.8e-3}$, and ultimate strain $\varepsilon_{u}=\num{7.1e-2}$.

\vspace{8mm}
\subsubsection*{Numerical Simulations} 

\noindent All finite element simulations were conducted using ABAQUS, in both 2D and 3D. For 2D simulations, we used four-node plane stress elements (element type: CPS4), while 3D simulations used eight-node brick elements with full integration (element type: C3D8). Individual cells were simulated in 2D, whereas full assemblies were simulated in 3D. The simulations were performed in two steps: first, an extension was applied via an imposed displacement, followed by the relaxation of elastic stresses. In the 2D simulations, one end of the cell was clamped during both steps. At the opposite end, a displacement $\delta$ was applied in the extension degree of freedom during the first step, with the two other degrees of freedom fixed. In the second step, all boundary conditions at the free end were released. For 3D simulations, which captured out-of-plane deformations, the ends of the parallel chains of cells were allowed to slide perpendicular to the direction of extension, mimicking the experimental setup. Isotropic elasto-plastic hardening, based on the von Mises yield criterion, was applied throughout all simulations. Mesh refinement was performed iteratively until convergence of the plastic displacement at equilibrium was reached. 

\subsubsection*{Rod model}  

\begin{figure}
  \centering
  \includegraphics[width=.8\textwidth]{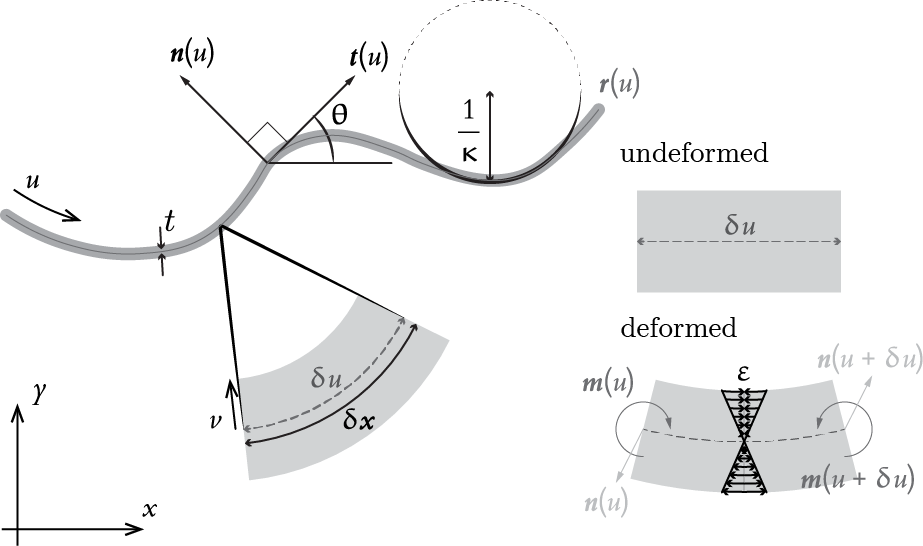}
  \caption{ \textbf{Rod model.} Sketch of the Kirchhoff rod centerline model and strain in undeformed and deformed configurations.}
  \label{fig:Methods1_rods}
\end{figure}

\noindent We describe the mechanics of our system using a Kirchhoff rod centerline model with an elasto-plastic material, considering an arbitrary distribution of initial curvature along the rod. The geometry for the Kirchhoff rod centerline model is illustrated in Fig.~\ref{fig:Methods1_rods}. The material coordinates $\bm{x}$ of a 2D rod with length $L$, depth $W$, and thickness $t$ are parameterized by the arclength $u$ and the normal offset $v$, as:
    \begin{align}
        \bm{x}(u,v)=\bm{r}(u)+v\bm{n}(u),
        \label{eq.pos}
    \end{align}
where $\bm{r}$ is the rod centerline position and $\bm{n}$ is the unit normal, as shown in Fig.~\ref{fig:Methods1_rods}. 
We assume that the cross-section remains planar, undistorted, and perpendicular to the centerline axis, ensuring that the material deformation is aligned with the centerline. Consequently, the material strain $\varepsilon$ is purely axial and can be defined as
    \begin{align}
        \varepsilon =  \frac{| \delta\bm{x} | - |\delta \bm{x}_0|}{|\delta \bm{x}_0|}.
        \label{eq.strain1}
    \end{align}
where the operator $|\bullet|$ represents the norm, and the subscript $(\bullet)_0$ denotes a variable for the undeformed reference state. The infinitesimal length $\delta \bm{x}$ of material points, as defined by equation~\ref{eq.pos}, is given by:
    \begin{align}
        \delta \bm{x} = \delta \bm{r} + v \delta \bm{n} = ( 1-\kappa v ) \bm{t} \delta u,
        \label{eq.len}
    \end{align}
where $\bm{t}$ is the unit tangent, $\kappa$ is the centerline curvature, and the transformation $\delta \bm{n} = - \kappa \bm{t} \delta u $ comes from the geometry of a plane curve.

\noindent By combining equations~\ref{eq.strain1} and~\ref{eq.len}, we derive the expression of the strain $\varepsilon$ of a material point located at $(u,v)$:
    \begin{align}
        \varepsilon (u,v) = v \frac{\kappa_0(u) - \kappa (u)}{1- v \kappa_0 (u)},
        \label{eq.strain2}
    \end{align}
where $\kappa$ represents the rod's current curvature, and $\kappa_0$ is the initial curvature.

\vspace{8mm}
\subsubsection*{Hinge model}  

\noindent We model the hinges as sections with a constant initial curvature $\kappa_0=1/R_0$ and assume that they deforms while maintaining a circular shape, with curvature $\kappa=(\pi-\alpha)/(\pi R_0)$, where $\alpha$ is the hinge angle. The strain in the hinge is then given by:  
\begin{equation}
    \varepsilon (v,\alpha) = -\frac{v\, \alpha}{\pi(R_0+v)},
    \label{eqn:strain2}
\end{equation}
assuming that the neutral axis (the axis of zero stretch) coincides with the centerline of the beam and that the cross-section remains orthogonal to the centerline. As the hinge opens ($\alpha: 0 \rightarrow \pi$), regions below and above the centerline experience tension and compression, respectively (see green and blue regions in Figure\,3c). Plastic regions appear when $|\varepsilon(v)| > \varepsilon_Y$, starting at the lower and upper surfaces. The coordinates $v_t$ and $v_c$, which delimit the plasticized zone in tension and compression respectively, are defined by $\varepsilon(v_t) = \varepsilon_Y$ and $\varepsilon(v_c) = -\varepsilon_Y$:
\begin{equation}
    v_{t,c} = \mp \frac{R_0 \pi \varepsilon_Y}{\alpha \pm \pi \varepsilon_Y}. \\
\end{equation}
where $v_{t}<0$ and $v_{c}>0$.

The plasticized region in tension (respectively in compression) appears when $v_t=-t/2$ (respectively $v_c=t/2$) at a critical hinge angle $\alpha_t$ (respectively $\alpha_c$), given by:
\begin{equation} 
    \alpha_{t,c} = \pi \varepsilon_Y \left(\frac{2 R_0}{t} \mp 1 \right),
\end{equation}

We compute the hinge moment:
\begin{equation}
    M(\alpha) = W \int^{t/2}_{-t/2} v \, \sigma(\varepsilon(v,\alpha)) dv
    \label{eqn:hingeMom}
\end{equation}
with $W$ is the beam width. We evaluate this moment in three distinct regimes: a fully elastic regime ($\alpha\leq \alpha_t$), a regime with a single plasticized region in tension ($\alpha_t < \alpha < \alpha _c $), and a third regime with plasticized regions in both tension and compression ($\alpha_c \leq \alpha$):
 
\begin{equation}
\frac{M(\alpha)}{W R^2_0/\pi} =
\begin{cases}
E\left(\log \left(\frac{1+\phi}{1-\phi}\right)-2\phi\right) \alpha, 
&  \alpha \leq \alpha_t \\
-\frac{1}{2}(E-E_{\text{p}}) \left( \left(1-\phi^2\right) (\alpha +\pi  \varepsilon _Y)-\frac{\alpha ^2}{\alpha +\pi  \varepsilon _Y} \right) - (E+E_{\text{p}})   \phi \alpha & \\
 -\alpha \left(E\log \left(\frac{\alpha }{(1+\phi) (\alpha +\pi  \varepsilon_Y)}\right)-E_{\text{p}} \log \left(\frac{\alpha}{(1-\phi) (\alpha +\pi  \varepsilon_Y)}\right)\right),  &  \alpha_t<\alpha\leq \alpha_c  \\
 -2 \alpha  E_{\text{p}} \phi - \alpha ^2  (E-E_{\text{p}}) \left(\pi  \varepsilon_Y  \left(1-\phi^2 \right)+ \frac{1}{2 (\alpha -\pi  \varepsilon_Y)}-\frac{1}{2 (\alpha +\pi  \varepsilon_Y)} \right) & \\
 +\alpha \left(E\log \left(\frac{\alpha +\pi  \varepsilon_Y}{\alpha -\pi  \varepsilon_Y}\right)-E_{\text{p}} \log \left(\frac{(1-\phi) (\alpha +\pi  \varepsilon_Y)}{(1+\phi) (\alpha -\pi  \varepsilon_Y)}\right)\right),
&  \alpha_c < \alpha 
\end{cases}
\end{equation}
where $\phi=t/(2 R_0)$ is the beam slenderness.
The asymptotic limit for the fully elastic regime and with two plasticized regions is shown in Fig.\,3c. In the limit of small curvatures ($\phi \ll 1$), the classical flexural modulus of a slender beam is recovered, as $\log \left(\frac{1+\phi}{1-\phi}\right)-2\phi \approx 2\phi^3/3$.

The moment $M(\alpha)$ can be approximated by a linear torsion spring model, $M(\alpha)\sim k(\alpha-\alpha_r)$ where the stiffness $k$ and reference angle $\alpha_r$ depend on whether the hinge angle has reached the critical angle $\alpha_c$. If $\alpha < \alpha_c$, the loading remains elastic, with $\alpha_r=0$ and the torsion spring stiffness given by $k_e = EWR_0b/\pi$, where $b = R_0 \log ((1+\phi)/(1-\phi))-t$. When $\alpha$ reaches the critical angle $\alpha_c$, the spring stiffness decreases significantly to $k_p = E_pWR_0b /\pi$ and $\alpha_r = -\pi\varepsilon_Y t^2(E/E_p-1)/(4R_0b)$. At this point, the hinge undergoes permanent deformation, with the plastic angle $\alpha_p = \alpha_m - M(\alpha_m)/k_e$ which depends on $\alpha_m$ and can be predicted by considering a linear elastic unloading with stiffness $k_e$.

\vspace{8mm}
\subsubsection*{Arm bending}  

\noindent The arms are modeled as Euler-Bernoulli beams. The deflection of the arm, $w$, is defined with respect to a coordinate system that is rotated by an angle $\alpha/2$ relative to the $x$-axis. This simplifies the system to a beam with ends clamped at an angle $-\beta$ relative to the rotated axis:
\begin{equation}
    EI w^{(4)} (x) = 0,
    \label{eq:EulerB}
\end{equation}
with $I=W t^3/12$. Integrating this equation with the boundary conditions $w(\pm L/2) = 0$ and $w'(\pm L/2) = -\tan{\beta}$, we obtain:
\begin{equation}
    w(x) = \frac{x}{2} \left(1-4\left(\frac{x}{L}\right)^2\right)\tan{\beta}.
\end{equation}
The bending energy stored in then given by:
\begin{equation}
    \frac{1}{2} EI \int^{L/2}_{-L/2} \left(w''(x)\right)^2 \, dx = E\frac{W t^3}{2 L} \tan^2{\beta}.
\end{equation}

\vspace{8mm}
\subsubsection*{Lagrangian formulation for symmetric cells}  

\noindent Geometry constrains the relationship between the two angles and the cell height. Assuming that arm shortening due to bending is small:
\begin{equation}
    \delta = 2 L \sin\left(\frac{\alpha}{2} + \beta\right) + 4 R_0 \frac{\pi}{\pi-\alpha} \cos\left(\frac{\alpha}{2}\right).
\end{equation}
The energy of the cell is expressed using a Lagrangian formulation:
 \begin{equation}
     \mathcal{L}(\alpha,\beta) = 2 \int^{\alpha}_0 M(\alpha') \, d\alpha' + E \frac{W t^3}{L} \tan^2{\beta} + F \left( \delta- \left(2 L \sin\left(\frac{\alpha}{2} + \beta\right) + 4 R_0 \frac{\pi}{\pi-\alpha} \cos\left(\frac{\alpha}{2}\right) \right) \right) .
 \end{equation}
where $F$ is a Lagrangian multiplier that corresponds to the axial force, with:
\begin{equation}
    \frac{d\mathcal{L}}{d\alpha} = 0, \label{eq:alpha}
\end{equation}

\begin{equation}
    \frac{d\mathcal{L}}{d\beta} = 0. \label{eq:beta}
\end{equation}
Eq.~(\ref{eq:alpha}) expresses the moment balance of the hinges:
\begin{equation}
    M(\alpha) = \frac{F}{2} \left(L \cos\left(\frac{\alpha}{2}+\beta\right) - \frac{2 \pi R_0}{\pi -\alpha} \left( \sin\left( \frac{\alpha}{2}\right) - \frac{2 \cos( \alpha/2) }{\pi - \alpha} \right) \right)
\end{equation}
Eq.~(\ref{eq:beta}) expresses the moment balance due to arm bending:
\begin{equation}
    F L \cos\left(\frac{\alpha}{2}+\beta\right) =  E \frac{W t^3}{L} \sec^2{\beta} \tan{\beta}
\end{equation}
We combine both equation to eliminate $F$ and obtain an implicit kinematic relation between $\alpha$ and $\beta$:
\begin{equation}
    M(\alpha) = E \frac{W t^3  \sec^2{\beta} \tan{\beta}}{2 L^2 \cos\left(\alpha/2 +\beta\right)} \left(L \cos\left(\frac{\alpha}{2}+\beta\right) - \frac{2 \pi R_0}{\pi - \alpha} \left( \sin\left( \frac{\alpha}{2}\right) - \frac{2 \cos( \alpha/2) }{\pi - \alpha} \right) \right) 
\end{equation}

\begin{figure}
  \centering
  \includegraphics[width=.8\textwidth]{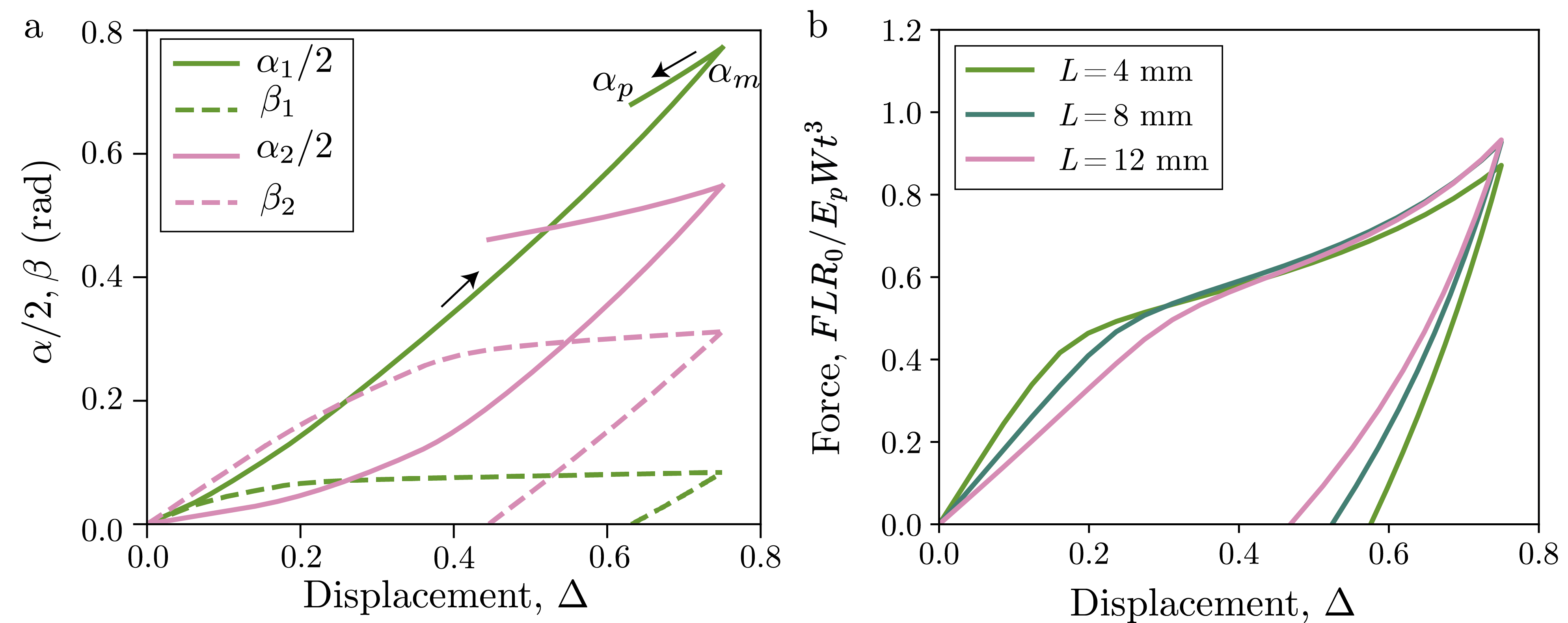}
  \caption{\textbf{Symmetric cells.} (\textbf{a}) Prediction of the angle $\alpha$ and $\beta$ plotted against normalized displacement $\Delta$. (\textbf{b}) Normalized force displacement curve.
}
  \label{fig:Methods1}
\end{figure}

This equation is solved numerically. In Fig.\,\ref{fig:Methods1}a, we show the variation of $\alpha$ and $\beta$ for two geometries as a function of the normalized strain parameter $\Delta=(\delta-\delta_0)/(2L + 2 \pi R_0-\delta_0)$, where $2L + 2 \pi R_0$ represents the curvilinear length of the rod. The maximum value of this parameter, $\Delta=1$, corresponds to fully straightened rod with no curvature. For the shorter arm ($L=4$\, mm), the arm hardly deforms ($\beta<0.1$), and the deformation is primarily concentrated in the hinge opening $\alpha$. Upon stress release, $\beta$ returns to zero, leaving permanent deformation localized in the hinge, where $\alpha$ relaxes from $\alpha_m$ to $\alpha_p$. In contrast the longer arm ($L=12$\, mm) experiences significant bending, resulting in a smaller hinge opening.

Fig.\,\ref{fig:Methods1}b illustrates the force variation as a function of stretch in cells for both geometries. The applied force is made dimensionless by $E_pWt^3/LR_0$ and is plotted against the normalized displacement $\Delta$. Curves for different arm lengths collapse in the plastic regime, further confirming the relevance of these gauges in our system.

Note that we can obtain an explicit expression in the limit where $R_0 \ll L$ and $\beta$ is small:
\begin{equation}
    \beta \approx - \tan^{-1}\left(\frac{2 L M(\alpha)}{E W t^3}\right).
\end{equation}

\vspace{8mm}
\subsubsection*{Origami length}  

\noindent The interplay between arm bending and hinge opening is reminiscent of the mechanics of a creased sheet,\cite{lechenault2014mechanical, dharmadasa2020formation} where the elastic origami length $\ell_e = \pi t^3 /(4 R_0 b)$\ determines whether cell extension is primarily due to hinge opening (when $L<\ell_e$) or arm bending (when $L>\ell_e$). When the plastic limit is surpassed (for $\alpha > \alpha_c$), the torsional stiffness of the hinge decreases dramatically.  At this point, hinge mechanics are governed by a new plastic origami length $\ell_{p}=(E/E_p)\ell_{e}$.
In our system, $\ell_{e}$ is approximately 50 times greater than $\ell_{p}$, meaning that once the plastic regime is reached, further extension is dominated by hinge opening (encoded by angle $\alpha$), while arm bending (encoded by angle $\beta$) ceases to increase (see Fig.\,\ref{fig:Methods1}a).

To achieve high curvature, one effective strategy is to use an asymmetrical cell with one arm long enough to keep its hinge in the elastic regime, while the shorter arm undergoes plastic deformation. The critical length $L^*$ separating these two regimes (as illustrated in the Inset in Fig.\,3b) depends on $\lambda$, $\varepsilon_Y$ and the elastic origami length $\ell_e$:
\begin{equation}
    L^*\sim 2 R_0 \sqrt{\frac{2 \phi (\lambda-1)}{\varepsilon_Y}}-\frac{\ell_e}{2}
\end{equation}

\vspace{8mm}
\noindent\textbf{Lagrangian formulation for asymmetric cells}  

\noindent The Lagrangian formulation of the energy of asymmetric cell (see Fig.2b) can be written as:
\begin{equation}
     \mathcal{L}_{C}(\alpha_1,\alpha_2,\gamma)=k_{el}\left( (\alpha_1 - \alpha_{1,p})^2+ (\alpha_2 - \alpha_{2,p})^2+(\alpha_1 +\gamma - \alpha_{1,p})^2+ (\alpha_2 -\gamma - \alpha_{2,p})^2 \right)+ \Lambda \, C(\alpha_1,\alpha_2,\gamma),
     \label{eq:lag}
 \end{equation}
 where $\Lambda$ is a Lagrange multiplier and $C(\alpha_1,\alpha_2,\gamma) = 0$ is a geometrical constraint that links the angles $\alpha_1$, $\alpha_2$ and the tilt angle $\gamma$ . Indeed, $H_1$ the height of the cell with length $L_1$ can be expressed as follows:
\begin{equation}
    H_1(\alpha_1,\gamma) = 2 L_1 \sin\left(\frac{\alpha_1}{2}\right) + \frac{2 \pi R_0 (2 \pi - \gamma - 2 \alpha_1)}{(\pi - \alpha_1)(\pi - \gamma - \alpha_1)}\cos\left(\frac{\alpha_1}{2}\right) - \frac{2 \pi R_0 + d (\pi -\gamma -\alpha_1)}{(\pi - \gamma - \alpha_1)}\sin\left(\frac{\gamma}{2}\right).
\end{equation}
 Similarly, $H_2$ the height of the cell of length $L_2$ is:
 \begin{equation}
    H_2(\alpha_2,\gamma) = 2 L_2 \sin\left(\frac{\alpha_2}{2}\right) + \frac{2 \pi R_0 (2 \pi + \gamma - 2 \alpha_2)}{(\pi - \alpha_2)(\pi + \gamma - \alpha_2)}\cos\left(\frac{\alpha_2}{2}\right) + \frac{2 \pi R_0 + d (\pi+\gamma -\alpha_2 )}{(\pi + \gamma - \alpha_2)}\sin\left(\frac{\gamma}{2}\right).
\end{equation}
Therefore $H_1(\alpha_1,\gamma)= H_2(\alpha_2,\gamma)$ and $C(\alpha_1,\alpha_2) = H_1(\alpha_1,\gamma) - H_2(\alpha_1,\gamma)$.
In the limit $R_0\ll (L_1,\, L_2)$ the geometrical constraint simplifies to  $C(\alpha_1,\alpha_2,\gamma)\approx 2L_1 \sin(\alpha_1/2)-2L_2 \sin(\alpha_2/2)$.

Numerical minimization of Eq.\ref{eq:lag} yields the equilibrium values of $\alpha_1$, $\alpha_2$ and the tilt angle $\gamma$ at equilibrium. The extension $\delta_{e}$ is calculated geometrically. The predicted tilt angle $\gamma$ as a function of arm lengths at constant force $F$ and uniform extension $\delta_e$ is shown in Fig.\,4c.

\vspace{8mm}
\subsubsection*{Generalized Kirchoff Model}  

\noindent The moment $\bm{m}$ of internal stresses $\sigma$ in the rod at a point $u$ is:
        \begin{align}
            \bm{m}(u) = - \int_A \sigma (u,v) \left( v- v_0(u) \right) dA,
            \label{eq.mom1}
        \end{align}
where $A$ is the cross section, $v_0$ is the neutral axis, and $\sigma$ is the stress. The neutral axis is determined by the condition that the rod is in pure bending so that there is no general extension or compression of the rod. The neutral axis can thus be found identically by the axial force $\bm{p} \cdot \bm{t}$ tending to zero so that the integral of stress in the cross-section should balance as:
        \begin{align}
             \int_A \sigma (u,v)  dA = \bm{p} \cdot \bm{t} \approx 0 
             \implies 
             \int_{-t/2-v_0(u)}^{t/2-v_0(u)} \int_{w} \sigma(u,v) dw dv = 0.
            \label{eq.neutral}
        \end{align}
The moment from internal stresses along the arclength $\bm{m}(u)$ for a plastic rod can then be solved given the undeformed curvature $\kappa_0(u)$, the deformed curvature $\kappa(u)$, and the bilinear constitutive model for stress $\sigma( \varepsilon)$ from equation~\ref{eqn:bilinear}. To determine the present geometry--defined by the curvature $\kappa(u)$--the derived moment is coupled to the Kirchhoff rod equations for force and moment balance along the arclength $u$, or respectively:
\begin{align}
    \frac{d \bm{p}}{d u} &= 0, \label{eqn.k1a}\\ 
    \frac{d \bm{m}}{d u} + \bm{t} \times \bm{p}  &= 0 , \label{eqn.k1b}
\end{align}
where $\bm{p}$ is the internal force. Noting that the moment is a function of the curvature $\bm{m}=f(\kappa(u),\kappa_0(u))$, that the curvature of a plane curve is defined $\kappa(u) = \tfrac{d \theta}{d u}$, and the tangent can be described $\tfrac{d\bm{r}}{du}=\bm{t}=\cos \theta \bm{e}_x + \sin \theta \bm{e}_y$, the 2D Kirchhoff equations and the rod centerline can be written as a system of six first order ordinary differential equations:
\begin{align}
    \frac{d p_x}{d u} = 0,&&  
    \frac{d p_y}{d u} = 0,&  \nonumber\\
    \frac{\partial \bm{m}}{\partial \kappa} \frac{d \kappa}{d u}+
    \frac{\partial \bm{m}}{\partial \kappa_0} \frac{d \kappa_0}{d u}+ 
    p_y \cos{\theta} - p_x \sin{\theta}  = 0,&&
    \frac{d\theta}{du} = \kappa,& \label{eqn.k2} \\ 
    \frac{d r_x}{du}=\cos \theta,&&  \frac{d r_y}{du}=\sin \theta.& \nonumber
\end{align}

Equations~\ref{eqn.k2} can be solved numerically given an initial rod geometry and boundary conditions during loading. For example, given the bar geometry of the symmetric hinge in Fig.\,5a under axial deformation, the initial curvature is the piecewise function:
\begin{equation*}
    \kappa_0(u) = 
    \begin{cases}
        -\frac{1}{R_0} & 0 \leq u \leq \frac{\pi}{2} R_0  \\
        0 & \frac{\pi}{2} R_0 < u \leq \frac{\pi}{2} R_0  + L \\
        \frac{1}{R_0} & \frac{\pi}{2} R_0  + L < u \leq \frac{3 \pi}{2} R_0 + L  \\
        0 & \frac{3 \pi}{2} R_0 + L  < u \leq \frac{3 \pi}{2} R_0 + 2 L   \\
        -\frac{1}{R_0} & \frac{3 \pi}{2} R_0 + 2L < u \leq 2 \pi R_0 + 2 L   
    \end{cases}
\end{equation*}
where $R_0$ is the radius of curvature of the hinge and $L$ is the length of the bar. The six constraints that imitate the constraints in experiments and finite element simulations of axial deformation and clamped ends are 
\begin{align*}
    r_x (0) &= 0,& r_y(0) &= 0,& \theta(0) &= \frac{\pi}{2},& \\
    r_x (2 \pi R_0 + 2 L ) &= 0,& r_y(2 \pi R_0 + 2 L ) &= \lambda 4 R_0 & \theta(2 \pi R_0 + 2 L ) &=\frac{\pi}{2},& 
\end{align*}
where $\lambda$ is the stretch factor. By solving equations~\ref{eqn.k2} with this geometry, we obtain the kinematics and force variation as a function of cell stretch, which we plot in Fig.\,5a. We note that the Kirchhoff rod model is in very good agreement with the finite element simulations.

\begin{figure}
  \centering
  \includegraphics[width=.8\textwidth]{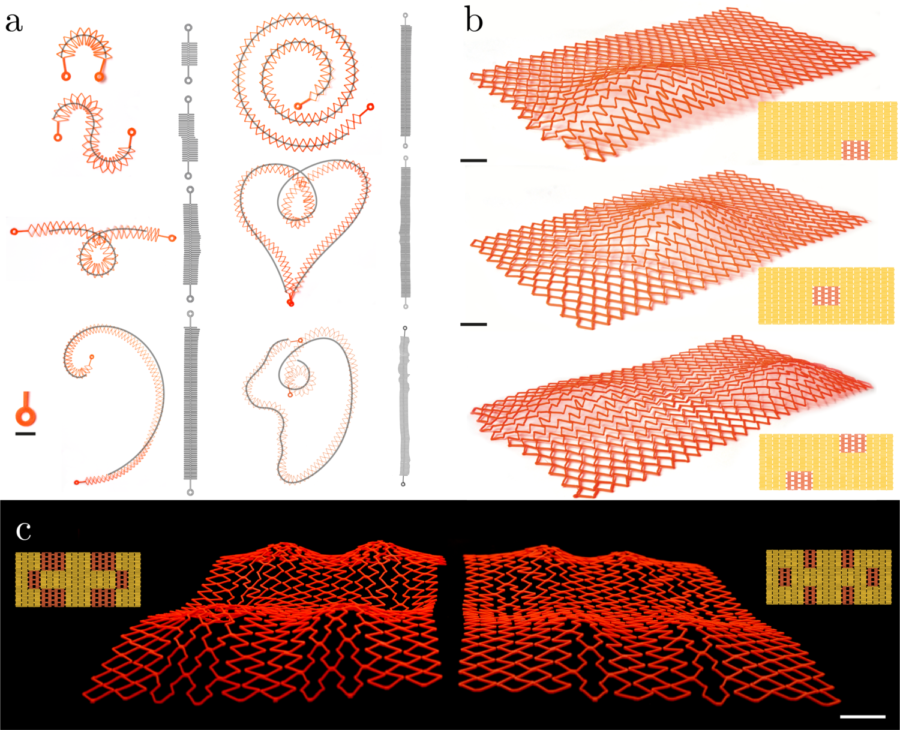}
  \caption{ \textbf{Control of curvature in 2D and 3D.} 
  (\textbf{a}) In 2D, chains of asymmetric cells are programmed to mach target trajectories (solid black line). All the shapes have the same extremities (scale bar: 7 mm). 
  (\textbf{b}) Localization of the Gauss curvature is programmed by embedding a two-by-three-cell patch into a uniform cell motif( scale bar: 20 mm). 
  (\textbf{c}) The magnitude of the curvature is controlled by varying the number of cells embedded in the uniform cell motif (scale bar: 20 mm).
}
  \label{fig:Methods3}
\end{figure}

\vspace{8mm}
\subsubsection*{Control of curvature in 2D and 3D and multistability}  

\begin{figure}
  \centering
  \includegraphics[width=.7\textwidth]{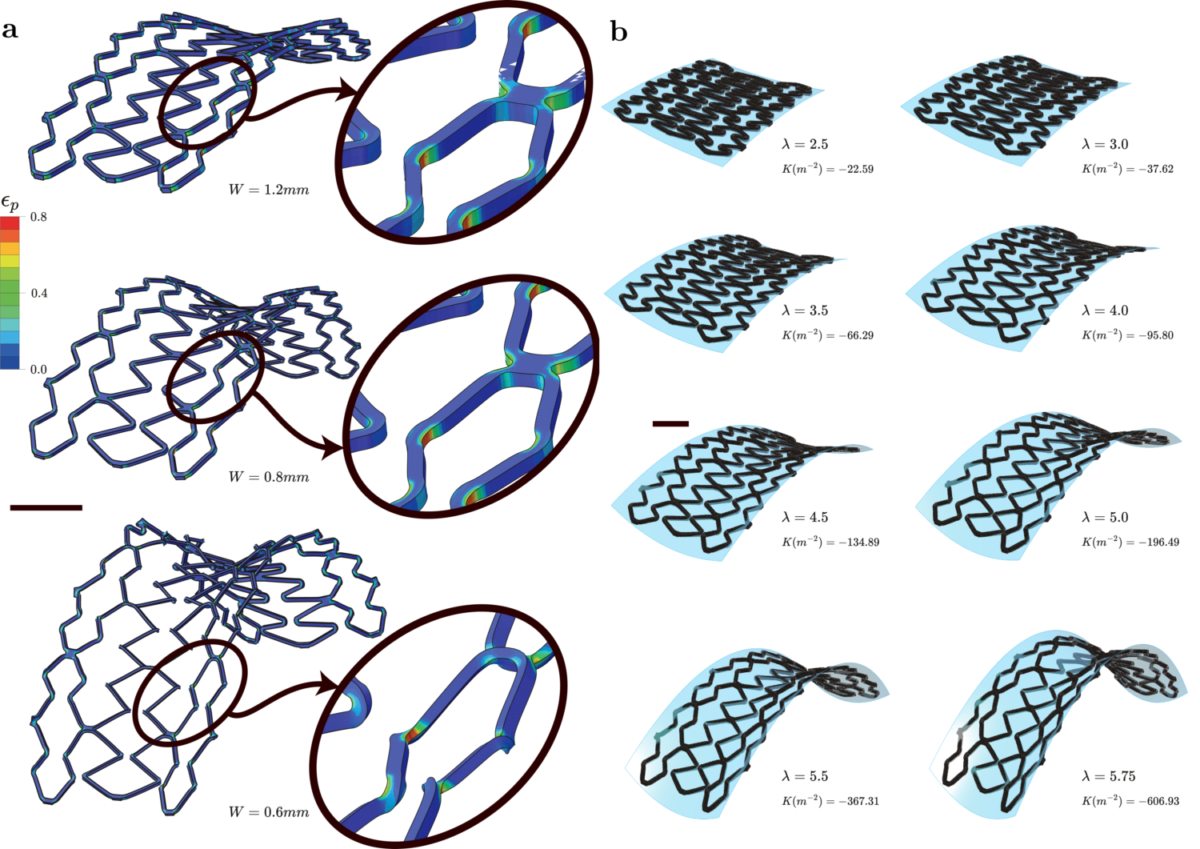}
  \caption{ \textbf{Kirigami effect and numerical equilibrium shapes} 
 (\textbf{a}) For a fixed geometry and stretch value $\lambda=5.5$, Gauss curvature increases as the plate thickness decreases. Out-of-plane rotations of the sections occur when $W<t$, where $W$ is the plate thickness and $t$ the beam thickness. FEM simulation show the regions of plastic strain. (\textbf{b}) Equilibrium 3D shapes from FEM simulations, Gauss curvature, and interpolated surfaces for $\lambda$ ranging from 2.5 to 5.75. For a given geometry, the Gauss curvature increases with the maximum stretch. Scale bars 20 mm. }
  \label{fig:kirigami}
\end{figure}

\begin{figure}
  \centering
  \includegraphics[width=.8\textwidth]{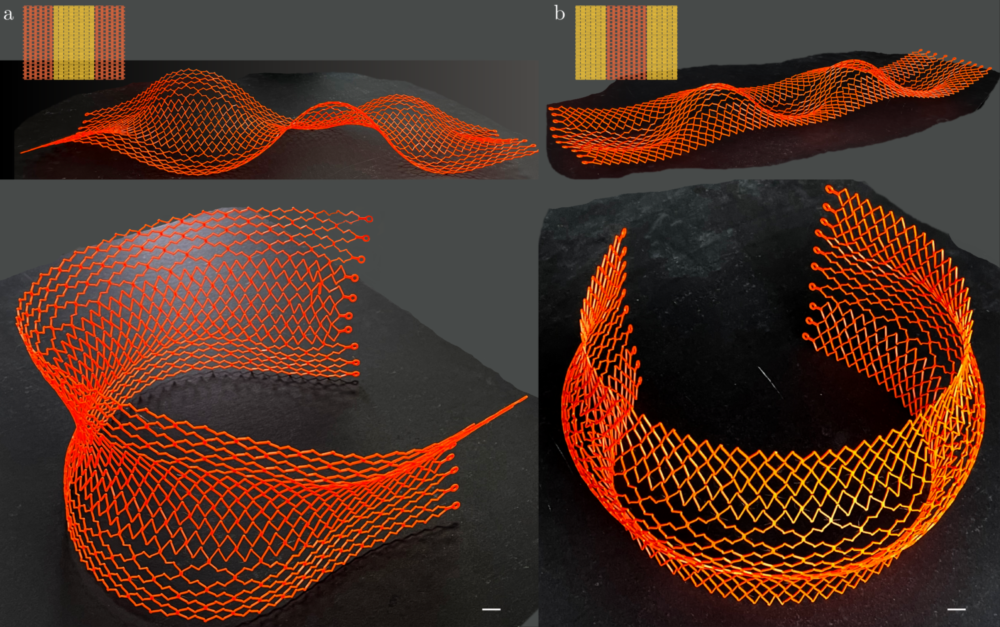}
  \caption{ \textbf{Multistable parallel chains} 
  Ten parallel chains after expansion, displaying negative Gauss curvature (\textbf{a}) and positive Gauss curvature (\textbf{b}). Two stable states are shown that are obtained by snapping one configuration to another. Scale bars 20 mm. }
  \label{fig:Methods6}
\end{figure}

\noindent The localization and magnitude of curvature can be precisely controlled in both 2D and 3D geometries. In 2D, solving the inverse problem allows the generation of target shapes with constant curvature, inflection points, self-intersections, and precise variations in curvature, e.g. Fibonacci and Archimedean spirals (see Fig.\,\ref{fig:Methods3}a). More complex shapes, such as hearts or faces, can also be produced. Increasing the curve's discretization enables the realization of sharp features. In 3D, Gauss curvature is achieved by programming metric changes (see Fig.\,\ref{fig:Methods3}b). The localization of the desired curvature is determined by the positioning of patches of heterogeneous cells, while the magnitude of the curvature can be adjusted by altering the internal arm length and extension of the heterogeneous cell patches (see Fig.\,\ref{fig:Methods3}c), or by increasing the maximum stretch, as demonstrated in Fig.\,\ref{fig:kirigami}b. For a given geometry and stretch value, Gauss curvature increases as the plate thickness decreases (see Fig.\,5d and Fig.\,\ref{fig:kirigami}a). The thinnest plates exhibit out-of-plane rotations at the hinges during extension, characteristic of kirigami deformation. Despite this, the structure maintains macroscopic Gauss curvature after unloading (see Fig.\,\ref{fig:kirigami}a). Longer systems exhibit multistability (see Fig.\,\ref{fig:Methods6}). After deformation, the curvature of the patches alternates direction, pointing upwards and downwards. The different stable shapes can be achieved by snapping from one configuration to another through the application of successive point forces. There is no preferential out-of-plane direction due to the anisotropy of the printing process, leading to an equal probability of obtaining samples where the first printed layer is on either the outside or inside of the deployed shell.

\end{document}